\documentclass[amsmath,amssymb,pra,aps,twocolumn,superscriptaddress,longbibliography]{revtex4-2}
\usepackage[colorlinks=true,citecolor=blue,linkcolor=red,urlcolor=blue]{hyperref}

\usepackage[english]{babel}
\usepackage{graphicx}% Include figure files
\usepackage{dcolumn}% Align table columns on decimal point
\usepackage{bm}% bold math
\usepackage{multirow}
\usepackage{xcolor}
\begin{document}

\title{Modular quantum extreme reservoir computing}

\author{Hon Wai Lau}
\affiliation{Okinawa Institute of Science and Technology Graduate University, Onna-son, Okinawa 904-0495, Japan}

\author{Aoi Hayashi}
\affiliation{School of Multidisciplinary Science, Department of Informatics,
SOKENDAI (the Graduate University for Advanced Studies),
2-1-2 Hitotsubashi, Chiyoda-ku, Tokyo 101-8430, Japan
}
\affiliation{Okinawa Institute of Science and Technology Graduate University, Onna-son, Okinawa 904-0495, Japan}
\affiliation{National Institute of Informatics, 2-1-2 Hitotsubashi, Chiyoda-ku, Tokyo, 101-8430, Japan}

\author{Akitada Sakurai}
\affiliation{Okinawa Institute of Science and Technology Graduate University, Onna-son, Okinawa 904-0495, Japan}

\author{William John Munro}
\affiliation{Okinawa Institute of Science and Technology Graduate University, Onna-son, Okinawa 904-0495, Japan}

\author{Kae Nemoto}
\affiliation{Okinawa Institute of Science and Technology Graduate University, Onna-son, Okinawa 904-0495, Japan}
\affiliation{National Institute of Informatics, 2-1-2 Hitotsubashi, Chiyoda-ku, Tokyo, 101-8430, Japan}

\date{\today}

\begin{abstract}
Quantum reservoir computing employs fixed quantum dynamics as a feature map for machine learning. 
Integrating multiple quantum reservoirs, however, raises a key question: 
how few inter-module connections are sufficient to match the performance of a single reservoir? 
To address this, we explicitly separate intra-module dynamics from inter-module couplings and systematically examine different connectivity schemes. 
We find that even a small number of well-placed connections between two modules can match single-reservoir accuracy, with simple one-to-one connections proving highly effective. 
Performance generally improves with increasing inter-module entanglement, and these correlations persist for both $ZZ$-coupled and random modular reservoirs. 
Extensions to three modules and evaluations across multiple datasets (MNIST, Fashion-MNIST, CIFAR-10) suggest that the modular architecture can be applied to diverse reservoir types and image-classification datasets. 
These results motivate modular quantum reservoir designs that align naturally with realistic hardware, such as two-dimensional quantum-chip layouts or networks of small integrated quantum systems.
\end{abstract}

%Use showkeys class option if keyword display desired
\maketitle

\section{Introduction}
\label{sec:introduction}

Quantum machine learning (QML) has gained significant attention for its potential to utilize quantum mechanics 
to perform faster and more accurate machine learning tasks 
\cite{biamonte_quantum_2017,cerezo_challenges_2022,huang_power_2021,schuld_quantum_2019,riste_demonstration_2017}.
Among various QML approaches, variational quantum algorithms (VQAs) 
\cite{cerezo_variational_2021,anschuetz_quantum_2022,skolik_quantum_2022,farhi_quantum_2014,tilly_variational_2022, mcclean_barren_2018,
bittel_training_2021, larocca_diagnosing_2022,cerezo_cost_2021,havlicek_supervised_2019} 
have shown promise for small problems by optimizing classical control of quantum gates.
However, issues related to trainability, such as barren plateaus, limit their scalability and practical applications 
\cite{mcclean_barren_2018,bittel_training_2021}.
Various solutions have been proposed to address these challenges 
\cite{larocca_diagnosing_2022,cerezo_cost_2021,havlicek_supervised_2019}, but no universal solution currently exists.

As an alternative to VQAs, quantum reservoir computing (QRC) is a promising framework that 
leverages the natural dynamics of quantum systems to process information in high-dimensional Hilbert spaces 
\cite{fujii_harnessing_2017,mujal_opportunities_2021,fujii_quantum_2021,gotting_exploring_2023, ghosh_quantum_2019,martinez-pena_dynamical_2021,
bravo_quantum_2022, xia_configured_2023,govia_nonlinear_2022,govia_quantum_2021,kora_frequency-_2024,dudas_quantum_2023,yasuda_quantum_2023,
kornjaca_large-scale_2024,innocenti_potential_2023,xiong_fundamental_2025}.
Typical QRC requires no training of the quantum reservoir itself, and the learning is confined to the classical part, therefore mitigating training problems in VQAs.
Recent experiments have demonstrated the potential effectiveness of QRC on near-term noisy intermediate-scale quantum devices \cite{chen_temporal_2020,bharti_noisy_2022},
with some evidence that noise may even enhance performance in certain settings \cite{fry_optimizing_2023,domingo_taking_2023,kora_statistical_2025}.
Theoretical works have also explored the expressivity and limitations of QRC \cite{xiong_fundamental_2025}.

Quantum extreme reservoir computing (QERC), as introduced in our previous works \cite{sakurai_quantum_2022,hayashi_impact_2023,sakurai_simple_2025},
is a quantum reservoir framework designed for supervised learning tasks such as image classification.
This approach uses quantum reservoirs as a fixed feature map to achieve high performance using physically realizable systems with only tens of qubits.
Early QERC investigations suggested that an effective reservoir should either behave like a random unitary or exhibit high complexity \cite{sakurai_quantum_2022}.
Subsequent studies found that simple Hamiltonians with distance-decaying $ZZ$ interactions followed by $X$ rotations 
are sufficient for strong performance \cite{sakurai_simple_2025}.
Further varying the distance decay suggested that neither all-to-all interactions nor nearest-neighbor interactions alone make effective reservoirs.

Motivated by the apparent sufficiency of finite-range interactions, 
we consider a modular architecture that explicitly isolates intra-module dynamics from inter-module quantum connections.
Such a modular separation naturally aligns with practical constraints encountered in experimental quantum platforms, including two-dimensional quantum chips and integrated small quantum systems.
In these realistic scenarios, fully connected interactions become challenging.
Therefore, a key question is whether such modular quantum reservoirs can match the performance of a single reservoir of equivalent total size, 
and how few inter-module connections are sufficient to achieve this performance.

\begin{figure*}[t]
\centering
\includegraphics[width=0.98\linewidth]{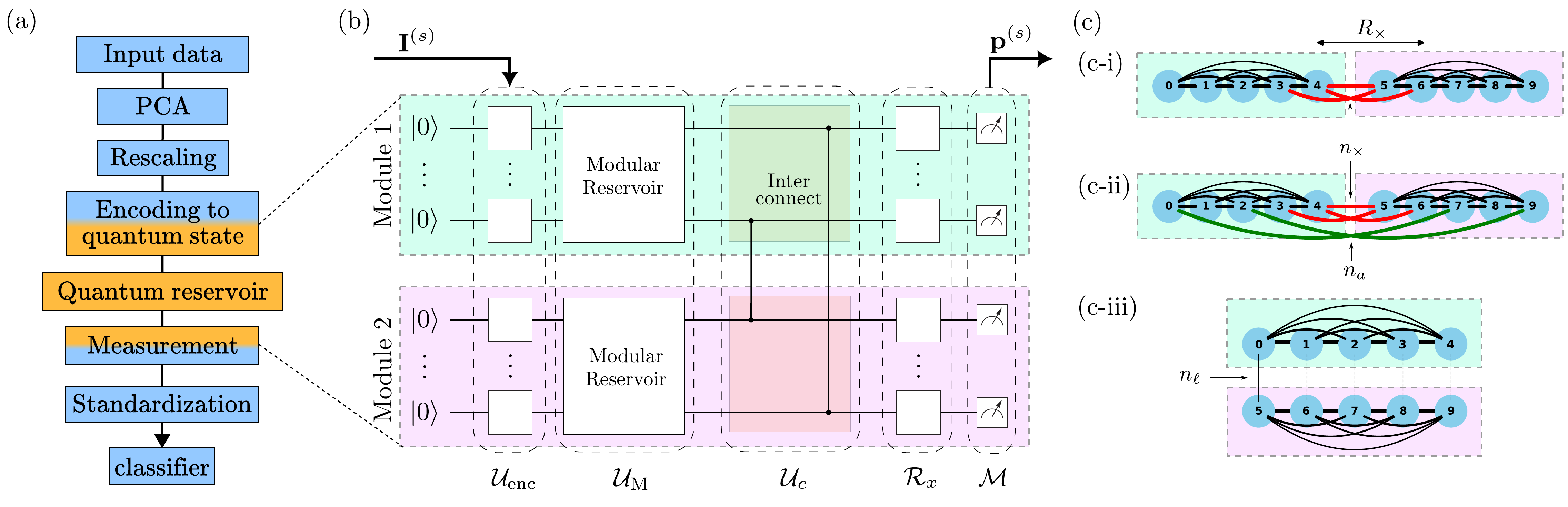}
\caption{
    Schematic illustration of MQERC and its architecture.
    (a) MQERC data flow involving classical pre-processing, a quantum reservoir, and a trainable classical linear classifier.
    (b) Quantum reservoir structure consisting of encoder $\mathcal{U}_{\mathrm{enc}}$ encoding processed classical data $\mathbf{I}^{(s)}$ into quantum states,
    module-level reservoirs $\mathcal{U}_{\mathrm{M}}$, 
    inter-module coupling $\mathcal{U}_c$, 
    single-qubit rotations $\mathcal{R}_x$, 
    and a measurement channel $\mathcal{M}$ resulting in the output $\mathbf{p}^{(s)}$.
    (c) Different inter-modular connectivity structures. 
    $R_\times$ is the boundary range, 
    $n_\times$ counts boundary connections, 
    $n_a$ counts additional arbitrary inter-module connections outside the $R_\times$ set,  
    $n_\ell$ counts the parallel connections.
    See the main text for details.
}
\label{fig:architecture}
\end{figure*}

To address these questions, we first introduce a modular QERC (MQERC) architecture in Section~\ref{sec:architecture}.
This allows us to investigate the performance contributions from modular reservoirs and the inter-module connections separately.
In Section~\ref{sec:single-chain}, we examine the effective range of a single-chain with $ZZ$ interactions to identify a good modular reservoir for subsequent analyses.
In Section~\ref{sec:two-modules}, we analyze boundary‑crossing, arbitrary, and one‑to‑one parallel connections to 
evaluate how different inter‑module connection schemes, coupling strengths, and the number of connections affect performance.
In Section~\ref{sec:entanglement}, we quantify bipartite entanglement between modules to examine how effectively MQERC utilizes the enlarged Hilbert space.
We further evaluate the generality of MQERC by repeating the analysis with random (CUE) modular reservoirs in Section~\ref{sec:random} and by extending to three modules in Section~\ref{sec:three-modules}. 
Finally, Section \ref{sec:discussion} summarizes the results with discussions.

\section{Architecture and Model}
\label{sec:architecture}

The modular QERC architecture (MQERC) aims to solve supervised classification tasks with image input:
given a labelled dataset $\{(\mathbf{x}^{(s)},y^{(s)})\}$ partitioned into training and test subsets, 
the objective is to maximize the test-set accuracy after learning from the training data.
The main focus of this paper is on the effects of the quantum connections between modular reservoirs on performance, quantified by test accuracy.

A schematic illustration of this architecture is depicted in Fig.~\ref{fig:architecture}, 
which consists of a classical pre-processor, a quantum reservoir, and a classifier using a one-layer classical neural network as shown in Fig.~\ref{fig:architecture}a.
To understand the effect of connections, 
we employ a general $m$-module structure for the quantum reservoir where the inter-module connections are explicitly separated from the details inside the modules, 
as illustrated in Fig.~\ref{fig:architecture}b.

For the pre-processor, PCA is used to select the important information in the input $\mathbf{x}^{(s)}$, 
and then the largest $2n$ rescaled PCA components $\mathbf{I}^{(s)}$ are encoded into the $n$ qubits 
through the encoder $\mathcal{U}_{\mathrm{enc}}(\mathbf{I}^{(s)})$ using $Y$ and $Z$ rotations.
The implementation is similar to \cite{sakurai_quantum_2022} and is detailed in Appendix A.

Next, the state is evolved under the action of the full quantum reservoir, which can be described by a unitary of the form $\mathcal{U} = \mathcal{R}_x \mathcal{U}_c \mathcal{U}_\mathrm{M}$.
As illustrated in Fig.~\ref{fig:architecture}b, the quantum reservoir architecture considered here consists of three components:
module-level reservoirs $\mathcal{U}_\mathrm{M}$, inter-modular connections $\mathcal{U}_c$, and single-qubit operations $\mathcal{R}_x$.

\begin{figure*}[tb]
\centering
\includegraphics[width=0.98\linewidth]{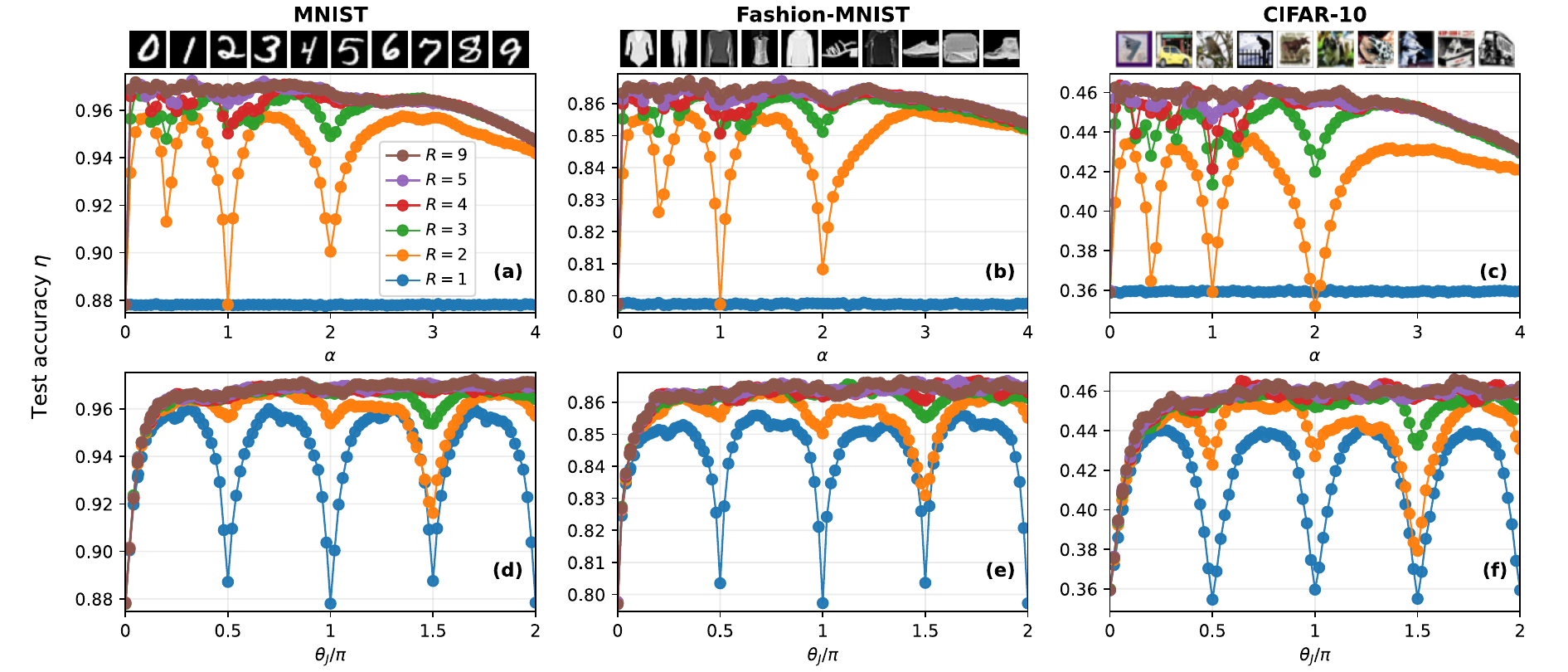}
\caption{
    Test accuracy $\eta$ for the image classification tasks using the (a) MNIST, (b) Fashion-MNIST, (c) CIFAR-10 datasets on a single 10-qubit chain vs interaction exponent $\alpha$
    for varying interaction range $R$ with $\theta_J=2\pi$ and $\theta_g=\pi/8$.
    Further we show for the (d) MNIST, (e) Fashion-MNIST, (f) CIFAR-10 datasets, $\eta$ vs $\theta_J$ at $\alpha=1.5$ for varying $R$.
}
\label{fig:mqerc_acc_alpha_R}
\end{figure*}

First, the module-level reservoirs $\mathcal{U}_\mathrm{M}$ comprise $m$ modules, specifically denoted by the sequence $[n^{(1)}, n^{(2)}, \ldots, n^{(m)}]$,
where each $n^{(\mu)}$ represents the number of qubits in the \(\mu\)-th module, so $n=\sum_{\mu=1}^{m} n^{(\mu)}$ is the total number of qubits in the whole system.
The modular reservoir for the $\mu$-th module is given by the unitary $\mathcal{U}^{(\mu)}$, 
meaning the overall action on the initial state by all modular reservoirs is described by $\mathcal{U}_\mathrm{M} = \bigotimes_{\mu=1}^{m} \mathcal{U}^{(\mu)}$.
There are many choices for the modular reservoir, which are usually based on physical systems, 
such as various Ising models and random unitaries \cite{sakurai_quantum_2022,hayashi_impact_2023}.
In particular, previous studies suggest that $ZZ$ interactions followed by $X$ rotations are sufficient to realize a high-performance reservoir \cite{sakurai_quantum_2022}.
Accordingly, the main reservoir we consider is described by the Hamiltonian
\begin{align}
H^{(\mu)} &= \hbar \sum_{i<j} J_{ij}^{(\mu)} Z_{i} Z_{j},
\label{eq:H_reservoir}
\end{align}
which gives the unitary $\mathcal{U}^{(\mu)} = \exp(-i \sum_{ij} \theta_{J,ij}^{(\mu)} Z_i Z_j)$ with $\theta_{J,ij}^{(\mu)} = J_{ij}^{(\mu)} t$.
The distance-dependent couplings within the same module are given by
$\theta_{J,ij}^{(\mu)} = \theta_J/|i-j|^\alpha,$
where $\theta_J$ is a control parameter of the reservoir, while $\alpha$ is the interaction exponent.
Typical regimes of $\alpha$ include $0 \lesssim \alpha \lesssim 3$ for trapped ions \cite{porras_effective_2004,britton_engineered_2012}, 
$\alpha=3$ for dipole-dipole interactions, 
and $\alpha=6$ for Van der Waals-type interactions \cite{bravo_quantum_2022,defenu_long-range_2023}.

Second, interactions between modules are introduced via $\mathcal{U}_c$ following the application of the local module operations $\mathcal{U}_\mathrm{M}$.
The inter-modular connections can take a different form from (\ref{eq:H_reservoir}), but here we choose the same $ZZ$ interactions for convenience.
This choice allows the use of the well-understood dynamics of $ZZ$ interactions and simplifies our analysis.
The inter-modular Hamiltonian has the form
\begin{align}
H_c &= \hbar \sum_{k,l} c_{kl} Z_k Z_l,
\end{align}
where the indices $k$ and $l$ now belong to different modules.
The corresponding unitary operation is given by $\mathcal{U}_c = \exp(-i \sum_{kl} \theta_{c,kl} Z_k Z_l)$ with $\theta_{c,kl} = c_{kl} t$.
We focus on the uniform case $\theta_{c,kl}=\theta_c$ for active connections, because this simplicity enables exhaustive evaluation of all configurations.
In particular, the interaction $Z_k Z_l$ with $\theta_c=\pi/4$ is an interesting regime, 
because it is equivalent to a controlled-$Z$ gate up to local operations, and in principle can create maximum entanglement between unentangled qubits.

Third, single-qubit operations $\mathcal{R}_x$ are applied before measurement to set the effective basis.
Since $\mathcal{U}_c$ is diagonal in the computational basis, applying $\mathcal{U}_c$ changes only phases.
Measuring directly in this basis would therefore leave outcome probabilities unchanged by $\mathcal{U}_c$, 
even if it generates inter-module entanglement.
Here, we use $\mathcal{R}_x = \prod_q R_{x}^{(q)}(\theta_g) = \prod_q \exp(-\tfrac{i}{2}\,\theta_g X_q)$ with $\theta_g=\pi/8$
to change the phase differences into amplitude differences.

Lastly, the final evolved state $|\psi_f\rangle$ is
\begin{align}
|\psi_f\rangle = \mathcal{R}_x \mathcal{U}_c \mathcal{U}_\mathrm{M} \mathcal{U}_{\mathrm{enc}}(\mathbf{I}^{(s)}) |0\rangle^{\otimes n},
\end{align}
which is then measured in the computational basis $\{|z\rangle\}$, where $z \in \{0,1\}^n$.
More specifically, the measurement channel $\mathcal{M}$ returns the probability $\mathcal{M}: |\psi_f\rangle \to \mathbf{p} \in \mathbb{R}^{2^n}$ with
\begin{align}
p_z = |\langle z|\psi_f\rangle|^2.
\end{align}
The probabilities $\{ \mathbf{p}^{(s)} \}$ for all training data are treated as features that are standardized and fed into a one-layer neural network classifier.
In this paper, we consider the true probabilities, and the effect of sampling was reported in \cite{hu_tackling_2023,sakurai_quantum_2025}.
The loss function is the cross-entropy and the training method is gradient descent using Adagrad, with learning rates of 0.05 and 0.002 for $n=10$ and $n=15$, respectively.
The details of the implementation and other training parameters are provided in Appendix A.

The main focus of this paper is on how the performance of MQERC on classification tasks is affected by the connections between modules.
More specifically, the performance is quantified by the smoothed test accuracy $\eta$, 
which is the fraction of test data classified correctly into the target class after training (see Appendix B).
The specific image classification tasks considered are MNIST, Fashion-MNIST, and CIFAR-10 
\cite{lecun_gradient-based_1998,krizhevsky_learning_2009,xiao_fashion-mnist_2017}, 
as detailed in Appendix C.

\section{Single chain with finite-range connections}
\label{sec:single-chain}

Before examining multi-module reservoirs and the effects of inter-module connections,
it is critical to analyze the internal structure of a single module ($m=1$).
In particular, we focus on the performance impact of reducing connectivity by limiting the interaction range within a single chain.
This analysis allows us to identify suitable modular reservoirs and 
establish a baseline for evaluating performance improvements in subsequent sections.

Specifically, we consider a single open chain of uniformly spaced qubits with interaction strength $\theta_{J,ij}$ defined by
\begin{equation}
\theta_{J,ij} = 
\begin{cases} 
\frac{\theta_J}{|i-j|^\alpha} & \text{for } 0 < |i-j| \leq R,  \\
0 & \text{otherwise}.
\end{cases}
\end{equation}
where $R$ is the maximum interaction range, and thereby controls the number of connections.

Fig.~\ref{fig:mqerc_acc_alpha_R}(a) shows the test accuracy $\eta$ for image classification of MNIST 
as a function of the interaction exponent $\alpha$ for different ranges $R$ in an $n=10$ qubit chain with $\theta_J=2\pi$.
The case $R=n-1=9$ allows all qubits to be connected directly and have high accuracies across a wide range of $\alpha=0.1 \sim 2$.
For $R \gtrsim 4$, the curves nearly overlap with the $R=9$ case.
As $R$ decreases, particularly for $2 \leq R \lesssim 4$, the performance dependence on $\alpha$ becomes more significant, 
with notable dips at specific values of $\alpha$ coinciding with phase‑commensurate points. 
For example, when $R=1$ with $\theta_J=2\pi$, the reservoir unitary $\mathcal{U}_\mathrm{M}$ is the identity.
Also when $R=2$ and $\alpha=1$, 
$\mathcal{U}_\mathrm{M}$ becomes the identity up to a global phase.
This happens less and less as $R$ increases.
The same qualitative behavior, including the dips, 
is observed for Fashion-MNIST and CIFAR-10 in Fig.~\ref{fig:mqerc_acc_alpha_R}(b,c), 
with differences only in the quantitative accuracies.
This is consistent with the general idea that high symmetry reduces reservoir efficiency.

To complement the above analysis of the dependence on $R$ and $\alpha$, 
Fig.~\ref{fig:mqerc_acc_alpha_R}(d-f) explores how the performance varies with the coupling strength $\theta_J$ at fixed $\alpha=1.5$.
For a small interaction range, $R=1$, the performance dips at specific $\theta_J$.
Increasing $R$ lessens these symmetry-related dips.
For the 10-qubit chain, $R \gtrsim 4$ is already similar to $R=9$ (all-to-all), 
and the highest performance is achieved near $\theta_J=2\pi$.
We also observe similar qualitative saturation for $n=8$ and $n=12$.
These dataset independence suggest that these behaviors originate from intrinsic reservoir properties rather than data-specific features.

Our results show that near-optimal performance can be achieved by a wide range of parameter choices, 
given a sufficiently large interaction range.
Accordingly, we adopt $\theta_J=2\pi$, $\theta_g=\pi/8$, and $\alpha=1.5$, consistent with previous results \cite{sakurai_simple_2025}.
With this setting, the 10-qubit single-chain performance with all-to-all connections is used as the baseline for subsequent comparison:
$97.04\%$, $86.43\%$, and $46.25\%$ for MNIST, Fashion-MNIST, and CIFAR-10, respectively.
The observation that $R>4$ gives no substantial improvement suggests that 
a $ZZ$-modular reservoir with $n^{(\mu)}=5$ and all-to-all connections is a good starting point, 
which will be used in the following sections.

\begin{figure*}[tb]
\centering
\includegraphics[width=0.98\linewidth]{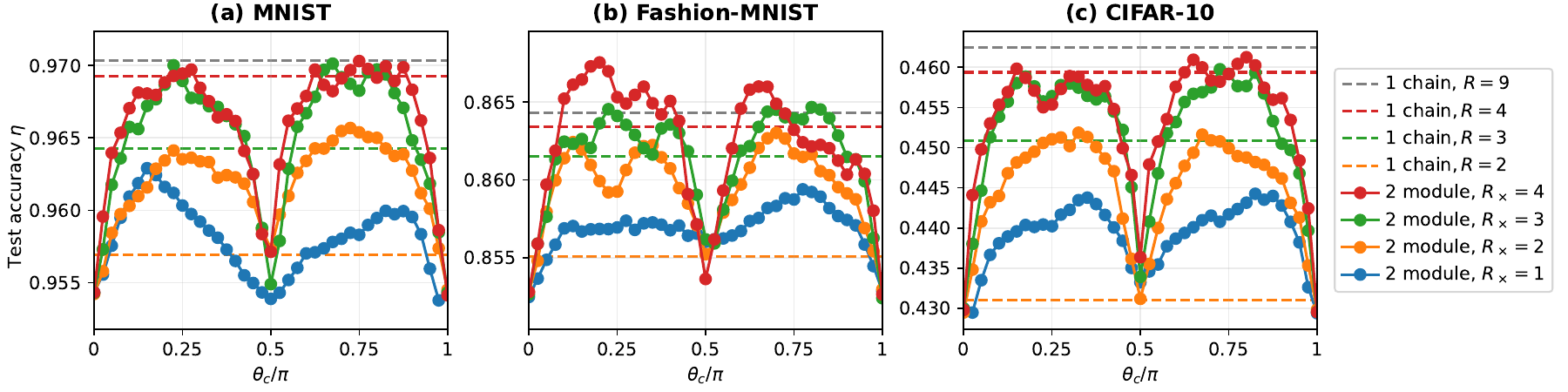}
\caption{
    Test accuracy $\eta$ vs inter-modular interaction strength $\theta_c$.
    Each curve represents a cross range $R_{\times}=1,2,3,4$ (corresponding to the number of cross connections $n_\times=1,3,6,10$).
    The horizontal dashed lines are 10-qubit single-chains with interaction range $R$ for comparison.
    Two modules with a [5,5]-module structure, with $\theta_J=2\pi$, $\theta_g=\pi/8$, and $\alpha=1.5$.
}
\label{fig:2module_thetac}
\end{figure*}

\definecolor{softgreen}{rgb}{0.1, 0.4, 0.1}  % Softer green
\definecolor{softblue}{rgb}{0.1, 0.1, 0.5}   % Softer blue

\begin{table*}[t]
\centering
\begin{minipage}{0.32\textwidth}
\centering
\begin{tabular}{|c|c|c|c|c|}
\hline
\textbf{(a)} &  & \textbf{$R_{\times}=0$} & \textbf{$R_{\times}=1$} & \textbf{$R_{\times}=2$} \\
 & \textbf{$n_a$} &(\textit{$n_\times=0$}) & (\textit{$n_\times=1$}) & (\textit{$n_\times=3$}) \\ \hline
 \multirow{3}{*}{$\alpha=1.5$} 
 & 0 & 0.9545 & \textcolor{softblue}{0.9602} & \textcolor{red}{0.9639} \\ \cline{2-5} 
 & 1 & \textcolor{softblue}{0.9627} & \textcolor{red}{0.9654} & - \\ \cline{2-5} 
 & 2 & \textcolor{red}{0.9664} & - & - \\ \hline
 \hline
 \multirow{3}{*}{$\alpha=2.0$} 
 & 0 & 0.9426 & \textcolor{softblue}{0.9491} & \textcolor{red}{0.9532} \\ \cline{2-5} 
 & 1 & \textcolor{softblue}{0.9507} & \textcolor{red}{0.9551} & - \\ \cline{2-5} 
 & 2 & \textcolor{red}{0.9569} & - & - \\ \hline
\end{tabular}
\end{minipage}
\hfill
\begin{minipage}{0.32\textwidth}
\centering
\begin{tabular}{|c|c|c|c|c|}
\hline
\textbf{(b)} &  & \textbf{$R_{\times}=0$} & \textbf{$R_{\times}=1$} & \textbf{$R_{\times}=2$} \\
 & \textbf{$n_a$} &(\textit{$n_\times=0$}) & (\textit{$n_\times=1$}) & (\textit{$n_\times=3$}) \\ \hline
 \multirow{3}{*}{$\alpha=1.5$} 
 & 0 & 0.8527 & \textcolor{softblue}{0.8567} & \textcolor{red}{0.8593} \\ \cline{2-5} 
 & 1 & \textcolor{softblue}{0.8600} & \textcolor{red}{0.8621} & - \\ \cline{2-5} 
 & 2 & \textcolor{red}{0.8650} & - & - \\ \hline
 \hline
 \multirow{3}{*}{$\alpha=2.0$} 
 & 0 & 0.8414 & \textcolor{softblue}{0.8454} & \textcolor{red}{0.8507} \\ \cline{2-5} 
 & 1 & \textcolor{softblue}{0.8494} & \textcolor{red}{0.8526} & - \\ \cline{2-5} 
 & 2 & \textcolor{red}{0.8557} & - & - \\ \hline
\end{tabular}
\end{minipage}
\hfill
\begin{minipage}{0.32\textwidth}
\centering
\begin{tabular}{|c|c|c|c|c|}
\hline
\textbf{(c)} &  & \textbf{$R_{\times}=0$} & \textbf{$R_{\times}=1$} & \textbf{$R_{\times}=2$} \\
 & \textbf{$n_a$} &(\textit{$n_\times=0$}) & (\textit{$n_\times=1$}) & (\textit{$n_\times=3$}) \\ \hline
 \multirow{3}{*}{$\alpha=1.5$} 
 & 0 & 0.4290 & \textcolor{softblue}{0.4407} & \textcolor{red}{0.4512} \\ \cline{2-5} 
 & 1 & \textcolor{softblue}{0.4478} & \textcolor{red}{0.4534} & - \\ \cline{2-5} 
 & 2 & \textcolor{red}{0.4554} & - & - \\ \hline
 \hline
 \multirow{3}{*}{$\alpha=2.0$} 
 & 0 & 0.4103 & \textcolor{softblue}{0.4186} & \textcolor{red}{0.4258} \\ \cline{2-5} 
 & 1 & \textcolor{softblue}{0.4235} & \textcolor{red}{0.4324} & - \\ \cline{2-5} 
 & 2 & \textcolor{red}{0.4360} & - & - \\ \hline
\end{tabular}
\end{minipage}
\caption{
    Highest test accuracy $\eta^*$ reported for each $(R_\times,n_a)$ and $\alpha = 1.5,2$.
    (a) MNIST, (b) Fashion-MNIST, and (c) CIFAR-10.
    $R_{\times}=0,1,2$ are cross ranges, corresponding to $n_{\times}=0,1,3$, respectively.
    $n_a$ is the number of additional arbitrary inter-module connections not counted by $n_\times$.
    Two modules with a [5,5]-module structure, with $\theta_J=2\pi$, $\theta_g=\pi/8$, and $\theta_c=\pi/4$.
}
\label{table:rcross-na}
\end{table*}

\section{Two modules}
\label{sec:two-modules}

We now focus on systems comprising two modular reservoirs ($m=2$) and their connectivity.
As suggested in the previous section, we use the [5,5]-module structure, which has the smallest effective modular reservoirs with all-to-all intra-module couplings ($R=4$), as the base modular reservoirs.
In this setting, we investigate the performance impact of connectivity schemes, the number of connections, and interaction strengths.
Three connectivity schemes that capture distinct physical constraints are considered:
(A) Boundary-crossing connections parameterized by a cross range $R_{\times}$ (equivalently $n_{\times}$ connections), 
(B) Arbitrary connections outside the $R_\times$ set, specified by a count $n_a$, and 
(C) One-to-one parallel connections specified by a count $n_{\ell}$.

\subsection{Boundary-crossing connections}
\label{sec:boundary-crossing-connections}

We first consider inter-module connections placed across the ends of two linear modules, as indicated by the red lines in Fig.~\ref{fig:architecture}c. 
This geometry allows a controlled comparison with the single-chain architecture with cutoff range $R=4$, since splitting the chain naturally yields two modules separated by a boundary.

Concretely, we introduce uniform boundary-crossing connections with a cross interaction range $R_\times$ defined by
\begin{equation}
\theta_{c,kl} = 
\begin{cases} 
\theta_c & \text{for } |k-l| \leq R_{\times}, \\
0 & \text{otherwise}.
\end{cases}
\end{equation}
where $k\in\{1,...,5\}$ and $l\in\{6,...,10\}$ label qubits in the first and second modules, respectively.
An example of such a configuration with $R_\times=2$ is illustrated in Fig.~\ref{fig:architecture}c-i.
Note that the number of boundary-crossing connections is $n_\times=0,1,3,6,10$ for $R_\times=0,1,2,3,4$ respectively.
$\theta_c$ is the uniform interaction strength for all inter-module connections.

Fig.~\ref{fig:2module_thetac} shows the test accuracy $\eta$ as a function of  $\theta_c$ for different $R_\times$.
For $R_\times > 0$, as $\theta_c$ increases from zero, the accuracy generally increases and reaches its highest values near, 
but not exactly at, $\theta_c \approx \pi/4$ or $\theta_c \approx 3\pi/4$.
At $\theta_c=\pi/2$, the accuracy drops to a value similar, but not identical, to that at $\theta_c=0$, 
even though both settings are unentangled as shown in Sec.~\ref{sec:entanglement}.
This behavior is consistent across all datasets.
Therefore, the result suggests that $\theta_c=\pi/4$ is indeed a good choice since the performance is quite close to optimal.

The horizontal dashed lines in Fig.~\ref{fig:2module_thetac} report the performance of a 10-qubit single-chain reservoir with different $R$.
For $R_{\times}=4$, the set of connections is the same as that of a 10-qubit single chain with $R=4$, 
except that the inter-module interaction strengths $\theta_{c,kl}$ differ.
For each dataset, there exists a $\theta_c$ that matches or slightly exceeds the performance of $R=4$, 
and some datasets have comparable performance at $\theta_c=\pi/4$.
Similar behavior is observed when comparing with the single chain with $R=9$, 
which has performance close to $R=4$.
These results suggest that MQERC can match a single-chain reservoir.

\subsection{Arbitrary connections}
\label{sec:arbitrary-connections}

In addition to the boundary-crossing connections, we consider arbitrary connections that can be placed between any two qubits across two modules.
It is interesting to see how much the non‑boundary‑limited inter‑module connections enhance performance compared with boundary-crossing $n_\times$.
In our [5,5]-module, there are at most $n^{(1)}n^{(2)}=25$ possible inter-module connections.
Throughout this subsection, we consider only $\theta_c=\pi/4$ and two qualitatively different interaction exponents $\alpha=1.5$ and $2$.

More precisely, the number of arbitrary connections $n_a$ we use is the number of connections between two modules that are not included in $R_\times$, so $0 \leq n_a \leq n^{(1)}n^{(2)} - n_\times$.
An example with $n_a=2$ is illustrated as green lines in Fig.~\ref{fig:architecture}c-ii, $R_\times=2$.
Because many configurations exist for a given $(R_\times,n_a)$, we report the highest test accuracy over all configurations, denoted by $\eta^*(R_{\times},n_a)$.
Note that $\eta^*(R_{\times},0) = \eta(R_{\times},0)$ by definition because there is only one configuration for a given $R_{\times}$.

First, we focus on $\alpha=1.5$ and MNIST with $n_a$ and $R_\times=0,1,2$ (equivalently $n_\times=0,1,3$).
For two completely separated modules with no inter-connections $(R_{\times}, n_a) = (0,0)$ as shown in the upper left of Table \ref{table:rcross-na}a, 
the highest test accuracy is $\eta^*(0,0)=95.45\%$.
Next, adding one boundary-crossing connection $R_{\times}=1$ increases the best performance to $96.02\%$, with $0.57\%$ difference.
On the other hand, adding one arbitrary connection $n_a=1$ increases the best performance to $96.27\%$, with $0.82\%$ difference,
which is $0.25\%$ higher than $R_{\times}=1$ above.
The more interesting result is that 2 arbitrary connections $\eta^*(0,2)$ can outperform 3 connections across the boundary $\eta^*(2,0)$.
Even $\eta^*(1,1)$ is strictly between $\eta^*(0,2)>\eta^*(1,1)>\eta^*(2,0)$.

These findings hold for other datasets and $\alpha=1.5,2$, suggesting that 
a few well-placed connections between two modules can achieve performance comparable to a higher number of connections across the boundary.

\subsection{One-to-one parallel connections}
\label{sec:parallel-connections}

\begin{figure}[tb]
\centering
\includegraphics[width=0.98\linewidth]{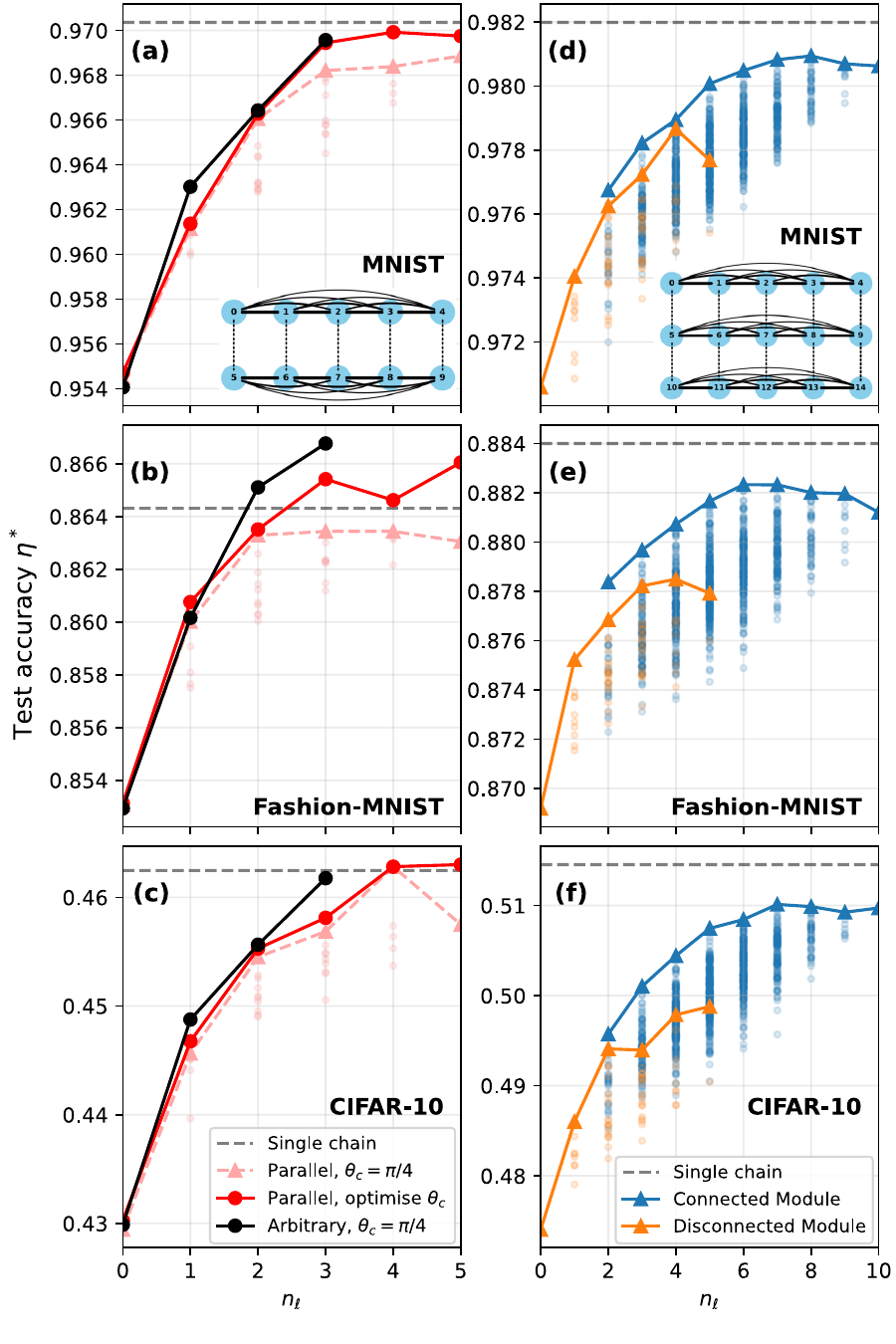}
\caption{
    Highest test accuracy $\eta^*$ over all possible configurations vs number of one-to-one parallel inter-modular connections $n_\ell$ for different datasets.
    (a-c) [5,5]-module with 32 possible parallel-connection configurations.
    Small pink circles: $\eta$ with $\theta_c=\pi/4$ for each configuration.
    Pink triangles: highest $\eta^*$ over configurations at fixed $n_\ell$ and $\theta_c=\pi/4$.
    Red circles: highest $\eta^*$ over configurations and $\theta_c$ at fixed $n_\ell$.
    Black circles: For arbitrary connections, positioned at $n_\ell = n_a$ on the x-axis, highest $\eta^*$ over configurations at fixed $n_a$ and $\theta_c=\pi/4$.
    (d-f) [5,5,5]-module with 1024 possible configurations.
    Connected: Both neighboring module pairs have at least one connection.
    Disconnected: Otherwise.
    Small circles and triangles: Same meaning as in (a-c).
    Dashed gray lines: Baselines of single chains with the same total qubit count.
    Insets: Possible inter-module connections are illustrated by dotted lines.
    $\theta_J=2\pi$, $\theta_g=\pi/8$, and $\alpha=1.5$.
    See the main text for details.
}
\label{fig:mqerc_parallel_acc}
\end{figure}

In addition to the boundary-crossing and arbitrary connections considered in the previous section, 
we now examine one-to-one parallel connections between modules, 
as illustrated by the example in Fig.~\ref{fig:architecture}c-iii.
This arrangement arises naturally for modular architectures in two dimensions, where inter-modular connections are more difficult to implement without intersections.

Specifically, we focus on two modules of equal size $n^{(1)}=n^{(2)}=n_0=5$, allowing at most $n_0$ parallel connections.
The set of possible parallel connections is given by $\{(i,i+n_0)\}$, 
as illustrated by dotted lines in the inset of Fig.~\ref{fig:mqerc_parallel_acc}(a-c).
Each configuration corresponds to a subset of these $n_0$ connections, giving a total of $2^{n_0}$ possible configurations.
For each number of active connections $n_\ell \in \{0,...,n_0\}$, we examine all $_{n_0}C_{n_\ell}$ possible configurations.

The test accuracy $\eta$ of each configuration with interaction strength $\theta_c=\pi/4$ is shown by small pink circles in Fig.~\ref{fig:mqerc_parallel_acc}(a-c).
The highest test accuracy $\eta^*$ for each $n_\ell$ is highlighted by pink triangles. 
Notably, in our data, the lowest test accuracy monotonically increases as more parallel connections are added.
These results clearly show that the performance increases with the number of parallel connections $n_\ell$ and saturates around $n_\ell=4$.
Further optimization over $\theta_c$, indicated by red circles, 
slightly enhances performance and can match the single-chain performance indicated by the dashed horizontal line.

For comparison, the performance for arbitrary connections, which are the broader class with less restricted configurations introduced in Sec.~\ref{sec:arbitrary-connections}, 
is included as black circles plotted against the number of arbitrary connections $n_a$.
While arbitrary connections typically achieve slightly better $\eta^*$, 
the number of potential configurations is significantly larger, and the search for an optimal configuration is challenging.

Overall, these results suggest that parallel connections already provide an effective and practical form of inter-module connectivity for MQERC. 
Moreover, even a single parallel quantum connection can result in a significant improvement in test accuracy.

\section{Entanglement and performance}
\label{sec:entanglement}

\begin{figure}[tb]
\centering
\includegraphics[width=0.98\linewidth]{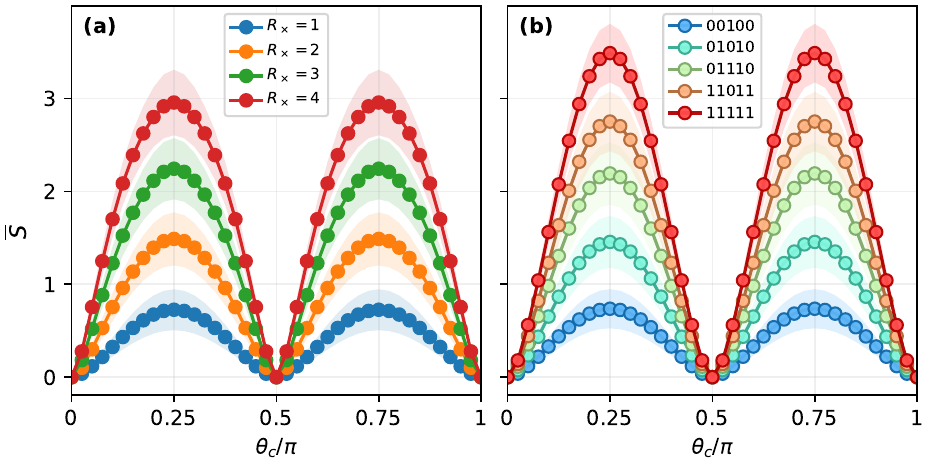}
\caption{
    Inter-module entanglement entropy $\overline{S}$ vs $\theta_c$ for MNIST.
    $\overline{S}$ is the entanglement entropy between the two modules in the final states,
    averaged over the test set, with a shaded area indicating one standard deviation.
    (a) Boundary-crossing connections $R_{\times}$ corresponding to Fig.~\ref{fig:2module_thetac}. 
    (b) One-to-one parallel connections $n_\ell$ corresponding to Fig.~\ref{fig:mqerc_parallel_acc}a, 
    with one specific configuration plotted for each $n_\ell$.
    A "1" in a configuration indicates the presence of a one-to-one parallel connection at the corresponding location for the [5,5]-module.
}
\label{fig:ee_vs_thetac}
\end{figure}

\begin{figure}[tb]
\centering
\includegraphics[width=0.98\linewidth]{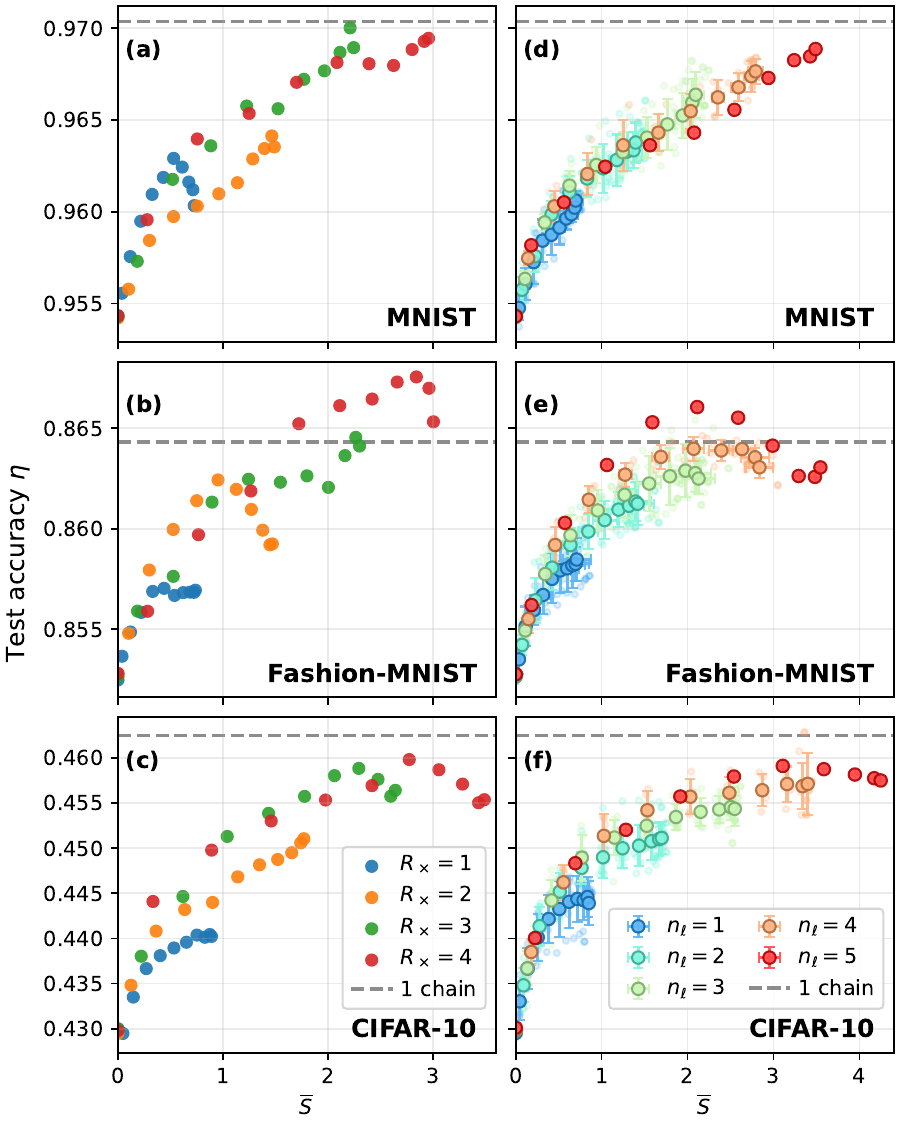}
\caption{
	Test accuracy $\eta(\theta_c)$ vs inter-module entanglement entropy $\overline{S}(\theta_c)$ with 11 uniformly spaced values in $\theta_c \in [0,\pi/4]$.
    (a-c) Boundary-crossing connections $R_{\times}$ corresponding to Fig.~\ref{fig:ee_vs_thetac}a.
    (d-f) One-to-one parallel connections $n_\ell$ corresponding to Fig.~\ref{fig:ee_vs_thetac}b.
    Small circles denote individual configurations.
    Large circles indicate the average $\eta$ and the average $\overline{S}$ over all configurations with the same $\theta_c$ and $n_\ell$.
    The error bars denote one standard deviation.
    Dashed gray lines: Baselines of single chains with the same total qubit count.
}
\label{fig:acc_vs_ee}
\end{figure}

The preceding sections demonstrated how the number of quantum connections and their interaction strength affect MQERC performance.
However, the exact relationship between the quantum correlations between modular reservoirs and the observed performance gains requires further investigation.
In this section, we quantify inter-module entanglement explicitly and examine its correlation with classification accuracy.

The two-module architecture provides a natural bipartition into two subsystems $A$ and $B$,
whose nonclassical correlations are captured by the bipartite entanglement entropy of the final state $|\psi_f^{(s)}\rangle$ defined in Sec.~\ref{sec:architecture}. 
We first compute the reduced density operator by tracing out subsystem $B$, 
\begin{equation}
\rho_A^{(s)} \;=\; \mathrm{Tr}_B\!\left[\,|\psi_f^{(s)}\rangle\langle\psi_f^{(s)}|\,\right],
\end{equation}
then calculate the von Neumann entanglement entropy
\begin{equation}
S^{(s)} \;=\; -\,\mathrm{Tr}\!\left[\rho_A^{(s)} \log_2 \rho_A^{(s)}\right].
\end{equation}
Since there are multiple inputs, we average over the test set with $N_{\mathrm{test}}$ data to obtain a representative measure of entanglement:
\begin{equation}
\overline{S}\;=\;\frac{1}{N_{\mathrm{test}}}\sum_{s=1}^{N_{\mathrm{test}}} S^{(s)} ,
\end{equation}
using the same test data as for the test accuracy $\eta$.
Smaller $\overline{S}$ indicates near-product states across the module partition, 
while larger $\overline{S}$ indicates stronger inter-module correlations.
This may provide a better feature map for the classical classifier and is therefore expected to improve performance.

Fig.~\ref{fig:ee_vs_thetac} illustrates how inter-module entanglement $\overline{S}$ grows with the coupling strength $\theta_c$ for two families of connections:
(i) boundary-crossing connections parameterized by $R_\times$ and
(ii) parallel connections parameterized by $n_\ell$.
In both cases, $\overline{S}$ initially increases smoothly from zero, 
attains a maximum at $\theta_c=\pi/4$, 
where performance is also near optimal in Fig.~\ref{fig:2module_thetac}, 
suggesting that strong entanglement is beneficial.
At the peak, both cases with one connection have an average $\overline{S} \approx 0.73$.
Generally, the theoretical bound is $\overline{S}\le 5$ per input state for the [5,5]-module, 
and an average $\overline{S}\approx 3.5$ is observed for parallel connections $n_\ell=5$ for MNIST. 
Subsequently, $\overline{S}$ decreases and vanishes at $\theta_c=\pi/2$.
Note that for each cross connection, 
$e^{-i\theta_c Z_k Z_l} = \cos \theta_c \mathbb{I} - i \sin \theta_c Z_k Z_l$.
At $\theta_c=\pi/2$, all connections are of the form $-i Z_k Z_l$, 
so the final states remain unentangled across the module bipartition.
The $\mathcal{R}_x$ rotations convert the local $Z$ phases from $\mathcal{U}_c$ into amplitude differences, 
so the final probability distribution and performance can differ from $\theta_c=0$.

The positive correlation between entanglement and performance in $\theta_c \in [0, \pi/4]$ is explicitly shown in Fig.~\ref{fig:acc_vs_ee}. 
The test accuracy $\eta(\theta_c)$ is plotted against the corresponding entanglement entropy $\overline{S}(\theta_c)$ for various $\theta_c$ values and connection schemes. 
For boundary-crossing connections shown in Fig.~\ref{fig:acc_vs_ee} (a-c), 
points for different $R_\times$ exhibit a positive correlation, 
falling approximately on a consistent curved trend line.
A similar positive relationship is observed for parallel connections in Fig.~\ref{fig:acc_vs_ee} (d-f).
Each small circle represents an individual configuration among $_5C_{n_\ell}$ possibilities at each given $n_\ell$.
These individual points align well along the same overall trend curve.
The trend is even clearer for the large circles, which represent the average $\eta$ and average $\overline{S}$ over different configurations at each $n_\ell$ and $\theta_c$.

The consistency of these observed positive trends across different datasets and connection schemes suggests that 
entanglement in modular reservoirs generally has a positive impact on performance.

\section{Random modular reservoirs}
\label{sec:random}

\begin{figure}[tb]
\centering
\includegraphics[width=0.98\linewidth]{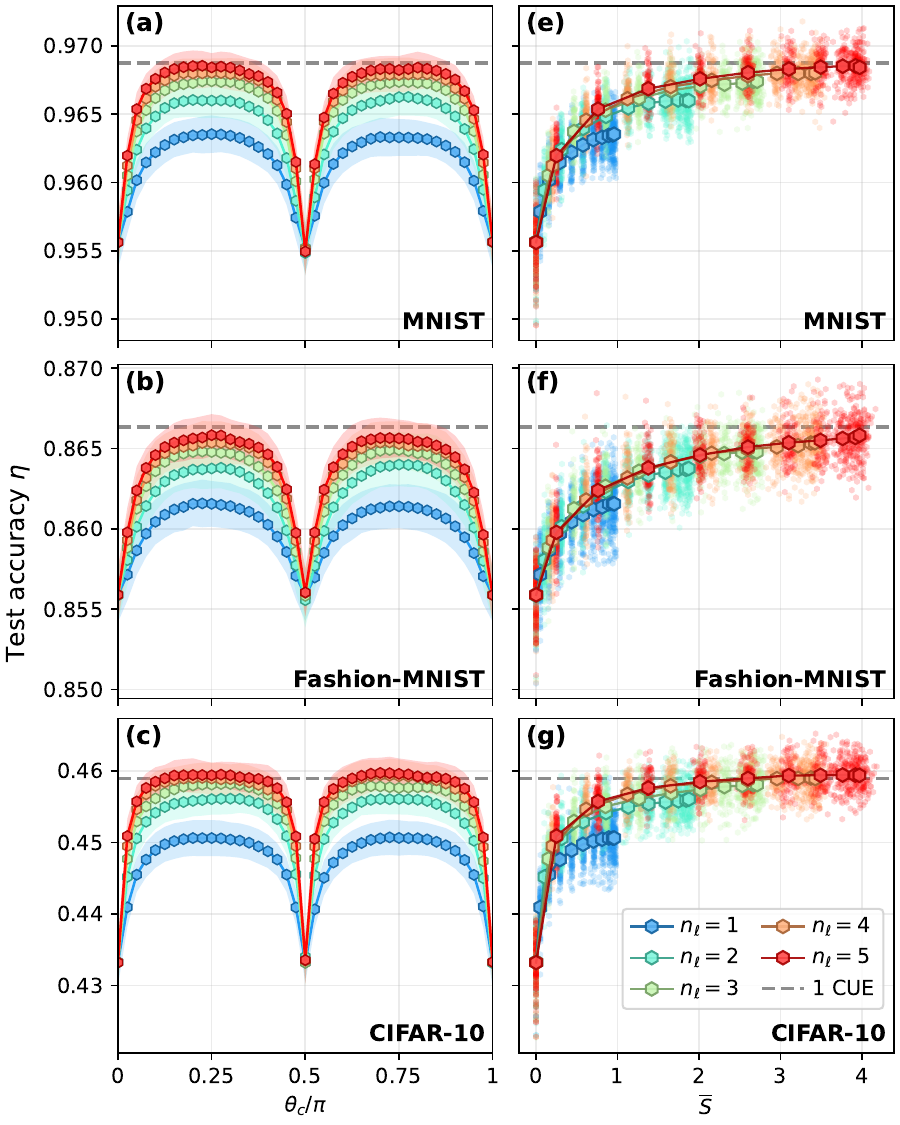}
\caption{
    Random modular reservoirs with one-to-one connections between two reservoirs.
    (a-c) Average test accuracy $\eta(\theta_c)$ vs $\theta_c$ plotted for different numbers of connections $n_\ell$.
    Each hexagon represents the average over 100 realizations of random modular reservoirs, with the shaded area indicating one standard deviation. 
    (d-f) Small hexagons plot $\eta(\theta_c)$ vs $\overline{S}(\theta_c)$ with 11 uniformly spaced values in $\theta_c \in [0,\pi/4]$ and $n_\ell$ across all realizations.
    Large hexagons indicate the average $\eta$ and the average $\overline{S}$ over all realizations with the same $\theta_c$ and $n_\ell$.
    Dashed gray lines: Baselines for a single random reservoir with the same total qubit count. 
    $\theta_g=\pi/8$.
}
\label{fig:cue}
\end{figure}

The modular architecture separates two main contributions to performance: 
the intrinsic properties of the modular reservoirs and the connections between them.
While the modular reservoirs examined so far are simple $ZZ$ reservoirs, 
earlier work indicated that their performance closely matches that of random reservoirs in single-module settings \cite{hayashi_impact_2023}.
Therefore, random modular reservoirs with one-to-one connections are good candidates for assessing the generality of the MQERC architecture.

Specifically, we sample random modular reservoirs from the Circular Unitary Ensemble (CUE).
For each realization, two modular reservoirs $\mathcal{U}^{(\mu)}$ ($\mu=1,2$) are independently drawn as random unitaries,
\begin{equation}
\mathcal{U}^{(\mu)} \sim \mathrm{CUE}(2^{n^{(\mu)}}),
\end{equation}
where $n^{(\mu)}$ is the number of qubits in module $\mu$.
The combined modular unitary is therefore $\mathcal{U}_\mathrm{M} = \mathcal{U}^{(1)} \otimes \mathcal{U}^{(2)}$.
Other architectural elements, including the subsequent inter-module connections, remain unchanged.
By Haar invariance of the CUE, any qubit relabeling within a module is equivalent statistically, 
so we can consider only $n_\ell$ one-to-one parallel connections without loss of generality.

Fig.~\ref{fig:cue}(a-c) shows the dependence of the average test accuracy $\eta$ on the inter-module coupling strength $\theta_c$ of a [5,5]-module, 
averaged over 100 random realizations.
Similar to Fig.~\ref{fig:2module_thetac}, the average $\eta$ increases from zero, 
and attains a maximum performance around $\theta_c=\pi/4$.
Nonetheless, the average over realizations here shows an even clearer smooth trend in the performance change with $\theta_c$.
Fig.~\ref{fig:cue} also shows the single random reservoir baselines with $\mathcal{U}_\mathrm{M} \sim \mathrm{CUE}(2^n)$ where $n = n^{(1)} + n^{(2)}$.
The interesting match of the performance suggests that 
two simple random modular reservoirs plus one-to-one connections can be as effective as a single large random reservoir.

With only one connection $n_\ell=1$ at $\theta_c=\pi/4$, 
the random modular reservoirs achieve accuracy improvements of 
$0.79\%$ ($95.56\%$ $\to$ $96.35\%$) for MNIST, 
$0.57\%$ ($85.59\%$ $\to$ $86.16\%$) for Fashion-MNIST, and 
$1.73\%$ ($43.33\%$ $\to$ $45.06\%$) for CIFAR-10.
The corresponding improvements for $ZZ$ modular reservoirs in Fig.~\ref{fig:mqerc_parallel_acc}(a-c) are 
$0.66\%$ ($95.45\%$ $\to$ $96.11\%$), 
$0.70\%$ ($85.30\%$ $\to$ $86.00\%$), and 
$1.62\%$ ($42.94\%$ $\to$ $44.56\%$) respectively,
which are comparable to typical random modular reservoirs.
Moreover, the optimal configurations and interaction strengths for the $ZZ$ modular reservoirs in Fig.~\ref{fig:acc_vs_ee}(d-f) 
can match and surpass slightly the average random modular reservoir performance.
This further supports the effectiveness of the $ZZ$ modular reservoirs.

Fig.~\ref{fig:cue}(d-f) illustrates the relationship between the test accuracy $\eta(\theta_c)$ and the entanglement entropy $\overline S(\theta_c)$.
Each small hexagon represents one realization together with a choice of $n_\ell$ and $\theta_c$.
All results approximately fall on a common curve increasing with entanglement.
This correlation is consistent with the trend observed for $ZZ$ reservoirs in Fig.~\ref{fig:acc_vs_ee}(d-f).
The large hexagons show the average $\eta(\theta_c)$ and average $\overline{S}(\theta_c)$, and the collapse of results for large $n_\ell$ is even clearer.

Collectively, these results show that the main conclusions drawn for $ZZ$ reservoirs can extend to random modular reservoirs.
That is, the number and strength of inter-module connections control both entanglement and accuracy. 
In addition, a few one-to-one connections are sufficient to recover the typical performance of a more complex single random reservoir, and the performance is positively correlated with inter-module entanglement.

\section{Three modules}
\label{sec:three-modules}

The performance improvement observed with one-to-one parallel connections in the previous sections suggests that 
this architecture may be sufficiently effective for MQERC.
To further explore this potential, we extend the analysis to a three-module architecture with parallel connections.

Specifically, we consider a modular reservoir with a [5,5,5]-module structure ($m=3$) and the same $ZZ$-type interactions as before.
As illustrated in the inset of Fig.~\ref{fig:mqerc_parallel_acc}(d), 
we consider only 10 parallel inter-module connections highlighted by dotted lines
(out of a maximum of $3\times 25=75$ possible inter-module connections).
Each pair of neighboring modules can support up to 5 parallel connections,
resulting in $0 \leq n_\ell \leq 10$ parallel connections and a total of $2^{10}=1024$ possible configurations. 
A configuration is connected if both neighboring module pairs have at least one connection, and disconnected otherwise.
The latter requires fewer resources to implement, 
so we examine it separately.

The test accuracy $\eta$ of each configuration with $\theta_c=\pi/4$ is shown in Fig.~\ref{fig:mqerc_parallel_acc} (d-f) as small circles.
The highest test accuracy $\eta^*$ for each number of parallel connections $n_\ell$ is highlighted by triangles.
$\eta^*$ generally increases with $n_\ell$ for different datasets, 
while the lowest test accuracy increases monotonically for all datasets, 
similar to Fig.~\ref{fig:mqerc_parallel_acc} (a-c).
For disconnected modules, the number of parallel connections is limited to at most five, and their performance saturates around $n_\ell = 4$.
In contrast, connected modules allow more parallel connections, 
and their performance saturates around $n_\ell = 7$ with accuracy approaching the single-chain baselines.

Note that at $n_\ell=0$, three independent modules in Fig.~\ref{fig:mqerc_parallel_acc}(d-f) outperform two independent modules in Fig.~\ref{fig:mqerc_parallel_acc}(a-c).
This suggests that combining multiple small independent quantum reservoirs can be a straightforward way to improve performance.
This improvement is partially attributed to the increased Hilbert-space dimension with more qubits, and the ability to process more principal components.
In particular, as demonstrated in Appendix~D, 
using 30 principal components ($n=15$) achieves higher accuracy than 20 principal components ($n=10$) without quantum reservoir.
Nevertheless, our results show that quantum connections remain essential for further enhancing overall performance.

\section{Conclusion and discussion}
\label{sec:discussion}
The modular architecture allows the explicit separation of modular reservoirs and quantum connections between them, 
and therefore allows a better understanding of the performance contributions from each component. 
In a single chain, finite-range $ZZ$ connections already match all-to-all performance, 
supporting modular designs of MQERC without full connections.
Examining boundary-crossing, arbitrary, and parallel connections shows that 
even a small number of well-placed connections can match single-chain performance.
In particular, one-to-one parallel connections provide a good trade-off between performance and layout simplicity.
Extending to three modules, parallel connections remain effective and approach single‑chain baselines.
We also show that performance correlates positively with inter‑module entanglement, 
and this correlation holds for both $ZZ$ reservoirs and random modular reservoirs, 
and across MNIST, Fashion‑MNIST, and CIFAR‑10.

These results suggest the possibility of simplifying the implementation of reservoir computing on quantum hardware \cite{chen_temporal_2020,krisnanda_experimental_2025}.
The modular architecture with non-crossing connections aligns naturally with layouts of two-dimensional systems such as quantum chips \cite{noauthor_ibm_2023}.
Similar modular structures have been considered in certain variational ansatzes to reduce entanglement and facilitate modularity in quantum circuits 
\cite{eddins_doubling_2022,peng_simulating_2020,fujii_deep_2022,diadamo_distributed_2021}.
In ion-based quantum computers with direct gates between any two qubits \cite{decross_computational_2024}, 
employing a small number of connections can still reduce computational time.
Alternatively, the integration of several small and simple quantum systems may be easier to implement in reservoir experiments, 
aligning more closely with distributed quantum computing \cite{barral_review_2024}.
In addition to image classification, temporal tasks may work well with a similar modular framework \cite{mujal_time-series_2023}.
Therefore, modular reservoirs have the potential to tackle complex machine learning tasks more effectively.

%In practice, this means that experimental platforms capable of implementing CUE-like dynamics locally (e.g.\ via random quenches) can match the performance of more elaborate engineered reservoirs with minimal overhead, provided that a few inter-module gates of controllable strength are available.
%These results suggest clearly that just one single quantum connection allows the improvement of the performance beyond the best of two independent modules. 
%Before adding any inter-connection, the best random reservoirs already perform 1\% higher than ZZ reservoirs. 

%\cite{barral_review_2024,king_performance_2020,van_meter_architecture_2008}

\section*{Appendix A: Classical processing and training}
\label{app:classical_processing}

This appendix provides the details of the non-quantum steps discussed in Section \ref{sec:architecture} as shown in Fig.~\ref{fig:architecture}(a).
Each step can be considered a data-processing step, and the last one is a classifier that is trained. 
The details of the classical components are similar to those in \cite{sakurai_quantum_2022}, but with minor modifications.

\subsection*{Notation and setup}
Let $\{(\mathbf{x}^{(s)}, \mathbf{y}^{(s)})\}_{s=1}^{N_{\mathrm{data}}}$ denote the inputs and one-hot targets, 
with $\mathbf{x}^{(s)}\in\mathbb{R}^{D}$, where $D$ is the data dimension. 
The dataset is split into training ($N_{\mathrm{train}}$) and testing ($N_{\mathrm{test}}$) sets. 
We use flattened images as inputs, as described in Appendix B. 
We consider a system with $n$ qubits, and in our simulations assume $2n \ll D \ll N_{\mathrm{train}}$.

\subsection*{PCA}
Principal component analysis (PCA) is used so that we can put important information into $n$ qubits. 
Only the training data are used to fit the PCA to prevent information from leaking into the test dataset.
Let $\boldsymbol{\mu} = \frac{1}{N_{\mathrm{train}}}\sum_{s\in\mathrm{train}}\mathbf{x}^{(s)}$ be the training mean 
and $\{\mathbf{v}_j\}_{j=1}^{D}$ the PCA orthonormal basis ordered by decreasing explained variance.
For each sample, we perform the projection
\begin{equation}
c_j^{(s)} \;=\; \mathbf{v}_j^{\top}\big(\mathbf{x}^{(s)}-\boldsymbol{\mu}\big), \qquad j=1,\dots,D.
\label{eq:pca-coeffs}
\end{equation}

\subsection*{Rescaling}
Each PCA coordinate is rescaled to $[0,1]$ so that we can cover a significant portion of the qubit surface.
Let $c_{j,\min} = \min \big[c_j^{(\mathrm{train})}\big]$ and $c_{j,\max} = \max \big[c_j^{(\mathrm{train})}\big]$, the rescaled data is
\begin{equation}
\tilde{I}_j^{(s)} = \frac{c_j^{(s)} - c_{j,\min}}{c_{j,\max} - c_{j,\min}},
\label{eq:minmax}
\end{equation}
which is then truncated to the range $[0,1]$ using clipping $I_j^{(s)} = \min(1,\max(0,\tilde{I}_j^{(s)}))$.

\subsection*{Encoding}
Data encoding can be powerful \cite{schuld_effect_2021}. 
Here we keep it simple to focus on the reservoir: 
we prepare $n$ qubits in $|0\rangle$ and encode the first $2n$ rescaled PCA components $\{I_j\}_{j=1}^{2n}$ using $Y$ and $Z$ qubit rotations.

Each data qubit $q \in \{1,...,n\}$ encodes two real numbers $(\theta_q,\phi_q) \leftarrow (\pi I_q,\pi I_{n+q})$.
Let $Y_q$ and $Z_q$ be Pauli operators acting on qubit $q$.
The single-qubit rotations are given by
\begin{equation}
R_y^{(q)}(\theta) = \exp\Big(-\frac{i}{2} \theta Y_q\Big),\quad
R_z^{(q)}(\phi) = \exp\Big(-\frac{i}{2} \phi Z_q\Big).
\label{eq:rot-defs}
\end{equation}
For qubit $q$, the encoded state prepared from $|0\rangle$ is
\begin{equation}
\begin{aligned}
|\psi_q\rangle
  &= R_z^{(q)}(\phi_q)\,R_y^{(q)}(\theta_q)\,|0\rangle \\
  &\propto \cos\frac{\theta_q}{2}\,|0\rangle
     + e^{i\phi_q}\sin\frac{\theta_q}{2}\,|1\rangle
\end{aligned}
\label{eq:single-qubit-state}
\end{equation}
up to a global phase.
The $n$-qubit encoded initial state $|\psi_{\mathrm{ini}}\rangle$ is the product state
\begin{equation}
|\psi_{\mathrm{ini}}\rangle = \bigotimes_{q=1}^{n}|\psi_q\rangle
 = \mathcal{U}_{\mathrm{enc}}(\mathbf{I}^{(s)}) |0\rangle^{\otimes n}.
\label{eq:encoded-register}
\end{equation}
where the encoding unitary is
\begin{equation}
\mathcal{U}_{\mathrm{enc}}(\mathbf{I}^{(s)}) = \prod_{q=1}^{n} R_z^{(q)}(\phi_q)\,R_y^{(q)}(\theta_q)
\end{equation}
The encoded state $|\psi_{\mathrm{ini}} (\mathbf{I}^{(s)})\rangle$ is then evolved under the unitary as described in the main text.

\subsection*{Standardization}
After the measurement, we have a probability vector $\mathbf{p}^{(s)} \in \mathbb{R}^{2^n}$ for data $s$.
Thus the quantum reservoir can be treated as a feature map from $2n$ to $2^n$ dimensions.
Let $\mu_z$ and $\sigma_z$ denote the componentwise mean and standard deviation of $\mathbf{p}$ computed over the training set. 
We standardize both training and test features as
\begin{equation}
\varphi_z^{(s)} \;=\; \frac{p_z^{(s)}-\mu_z}{\sigma_z},\qquad z\in\{0,1\}^n.
\label{eq:zscore}
\end{equation}
Note that this standardization method is commonly used in machine learning and differs from previous works \cite{sakurai_quantum_2022,hayashi_impact_2023}.

\subsection*{One-layer classifier and training}
Let $m_{\mathrm{class}}$ be the number of classes. The linear logits and softmax read
\begin{align}
u_c^{(s)} &= \sum_{z\in\{0,1\}^n} \varphi_z^{(s)}\,w_{zc}+b_c,\qquad c=1,\ldots,m_{\mathrm{class}}, \\
\hat{y}_c^{(s)} &= \frac{\exp(u_c^{(s)})}{\sum_{r=1}^{m_{\mathrm{class}}}\exp(u_r^{(s)})}.
\end{align}
With one-hot target $\mathbf{y}^{(s)}$, the per-sample cross-entropy loss is
\begin{equation}
L^{(s)} \;=\; -\sum_{c=1}^{m_{\mathrm{class}}} y_c^{(s)} \log \hat{y}_c^{(s)}.
\end{equation}
The parameters $(W,\mathbf{b})$ are optimized using Adagrad.
We use TensorFlow to implement the classifier and training with a minibatch size fixed at 100.

\section*{Appendix B: Test accuracy}
\label{app:test_accuracy}
The performance reported in the main text is the smoothed test accuracy $\eta$. 
Since we are exploring the effect of the connections on performance, smoothing the test accuracy reduces fluctuations and reveals a clearer relation. 
For each parameter setting, we use $N_\mathrm{run}$ training realizations, each trained for $E$ epochs, 
and average the test accuracy over the last $W$ epochs within each run, then average across realizations:
\[
\eta \;=\; \frac{1}{N_\mathrm{run}}\sum_{u=1}^{N_\mathrm{run}}\left(\frac{1}{W}\sum_{e=E-W+1}^{E} a_{u,e}\right),
\]
where $a_{u,e}$ is the test accuracy at epoch $e$ for realization $u$. 
We set $(N_\mathrm{run},E,W)=(3,\,100,\,20)$.

\section*{Appendix C: Dataset}
\label{app:datasets}

The data are provided by TensorFlow Datasets. Here, we describe the datasets we used and their properties. 

\textbf{MNIST:}  
The MNIST handwritten digit dataset \cite{lecun_gradient-based_1998} contains 
$60000$ training and $10000$ test grayscale images of size $28\times 28$ across $10$ classes.
Each class is a handwritten digit from 0 to 9 in two dimensions.
Suppose $a$ and $b$ are the two-dimensional indices.
Each image $x_{ab}$ is flattened to form the input $x_{ab} \to x_j \in \mathbb{R}^{784}$.
 
\textbf{Fashion-MNIST:}
Fashion-MNIST \cite{krizhevsky_learning_2009} mirrors the structure of MNIST with 
$60000$ training and $10000$ test grayscale images of size $28\times 28$ across $10$ classes.
Each class is a fashion category, including t-shirt, trouser, dress, sandal, etc. 
We apply the same flattening as MNIST.

\textbf{CIFAR-10:}
CIFAR-10 \cite{xiao_fashion-mnist_2017} contains 
$50000$ training and $10000$ test color images of size $32\times 32$ across $10$ classes.
Each class corresponds to a different object, including airplanes, cats, dogs, ships, etc.
A color image $x_{abc}$ is indexed by two-dimensional indices $(a,b)$ with three color channels $c$.
It is flattened to form the input $x_{abc} \to x_j \in \mathbb{R}^{3072}$.

\section*{Appendix D: PCA-only performance}
\label{app:pca_baselines}

\begin{table}[t]
\centering
\caption{Test accuracy (\%) of PCA-only reference values without the quantum reservoir.}
\label{tab:pca-only}
\begin{ruledtabular}
\begin{tabular}{lccc}
Dataset & PCA 20 & PCA 30 & PCA 784 \\
\hline
MNIST         & 86.87\% & 89.12\% & 92.47\% \\
Fashion-MNIST & 79.29\% & 80.70\% & 83.92\% \\
CIFAR-10      & 35.36\% & 36.33\% & 41.07\% \\
\end{tabular}
\end{ruledtabular}
\end{table}

The classical PCA-only performance obtained by sending data to bypass the quantum reservoir are calculated here. 
The rescaled principal components (PCA) are fed directly into the standardization layer, 
and then into the one-layer classifier described in Appendix A (see Fig.~\ref{fig:architecture}a for the architecture). 
Tab.~\ref{tab:pca-only} reports three different PCA settings: 20 components (matching the $n=10$ qubit setting that encodes two components per qubit), 
30 components (matching the $n=15$ qubit setting), and 784 components. 
All other preprocessing, splits, and training settings follow Appendix~A.

\section*{Acknowledgements}
We thank Henry L. Nourse and H. Yamada for valuable discussions and comments on this project.
This work is supported in part by the MEXT Quantum Leap Flagship Program (MEXT Q-LEAP) under 
Grant No. JPMXS0118069605, 
COI-NEXT under Grant No. JPMJPF2221, 
the Japanese Cross-ministerial Strategic Innovation Promotion Program (SIP) under Grant No. JPJ012367, 
and the JSPS KAKENHI Grant No. 21H04880.

%\section*{Author contributions statement}
%H.L. ....

%\bibliographystyle{apsrev4-2}
\bibliography{ref}
%\bibliography{all}

\end{document}